\documentstyle[rawfonts,nassau,pslatex]{article}  

\frompage{000} \topage{000}                                              
\def\rightmark{Particle Correlations with the PHENIX Experiment} 
\def\leftmark{S. C. Johnson}

\newcommand {\ppbar}    {\mbox{$p\overline{p}$}}

\newcommand {\Rinv}     {\mbox{$R_{inv}$}}

\newcommand {\Rs}       {\mbox{$R_{side}$}}
\newcommand {\Ro}       {\mbox{$R_{out}$}}
\newcommand {\Rl}       {\mbox{$R_{long}$}}

\newcommand {\qs}       {\mbox{$q_{side}$}}
\newcommand {\qo}       {\mbox{$q_{out}$}}
\newcommand {\ql}       {\mbox{$q_{long}$}}

\newcommand {\kT}       {\mbox{$k_T$}}
\newcommand {\mT}       {\mbox{$m_T$}}

\newcommand {\snn}      {\mbox{$\sqrt{s_{NN}}$}}

\newcommand {\RoPCMS}   {\mbox{$R_{out}^{PCMS}$}}
\newcommand {\RoLCMS}   {\mbox{$R_{out}^{LCMS}$}}
\newcommand {\RoTRUE}   {\mbox{$R_{out}^{TRUE}$}}

\title{Particle Correlations
with the PHENIX Experiment}
\authors{
{Stephen C. Johnson}\\[2.812mm]
{\normalsize
Physics and Advanced Technologies Directorate \\
Lawrence Livermore National Lab \\ 
Livermore, CA 94550\\[0.2ex] 
}}
 
\abstract{Results of identical pion correlations from the first year of
data collection with the PHENIX detector at RHIC (\snn=130GeV) are
presented.  PHENIX has good particle identification using an
electromagnetic calorimeter for timing, leading to identified pions
from .2 to 1 GeV/c.
This extends the range of previously measured correlation radii at
this energy to
$\langle k_T \rangle$=633MeV/c.  The beam energy dependence of the HBT
radii are studied in depth and no significant dependence of the
transverse radii is present.  The longitudinal correlation length has
a moderate energy dependence.  Furthermore, theoretical predictions of
$\Ro/\Rs$ severely underpredict the measured ratio, which is
consistent with unity for all $k_T$.  
The implications of these results are considered.}
 
\begin{document}
 
\maketitle

\section{Theoretical Considerations}\label{intro}

\subsection{The Physics of GGLP}

A distortion of the two particle probability distribution by the
influence of Bose-Einstein statistics was first noted by the quartet
of physicists Goldhaber, Goldhaber, Lee, and Pais in \ppbar~collisions
in 1960 \cite{gol60}.  This observable was subsequently adopted in a
variety of systems as a technique to probe the space-time history of
the collision region.  For a recent review see \cite{wie99}.

The normalized two particle probability distribution,
$P(p_1,p_2)/P(p_1)P(p_2)$, is given by the ratio of the two particle
probability distribution to the square of the single particle
distributions.  This function is defined as the `correlation function'
and, for a static source with no final state interactions, is related
to a Fourier transform of the source spatial distribution ($\rho(r)$)
with respect to the momentum difference, $q$ ,
$C_2\approx1+{ \tilde{\rho}(q)}^2$.  A more general expression for the
correlation function is:

\begin{equation}
C(\vec{q},\vec{k}) = 1 + 
\frac{|\int d^4xS(x,p)e^{iq\cdot x}|^{2}}{|\int d^4xS(x,p)|^{2}}
\label{eq:corr}
\end{equation}
where $\vec{k}$ is the momentum average of the pair.

Due to experimental statistical limits, it has to date been
practically impossible to study the full six dimensional correlation
function.  Instead, in most cases we take advantage of the event
averaged impact parameter to reduce our nominally azimuthally
asymmetric source into a symmetric one \footnote{see
\cite{lisaAzimuth} for a very interesting counter example.}:
$(\vec{q},\vec{k})\rightarrow(\vec{q},k_T,y)$ where $k_T$ is the
component of $\vec{k}$ perpendicular to the beam.  Historically,
experimentalists have plotted $C_2(\vec{q})$ while cutting on $k_T$
and $y$.

\subsection{A Note on Frames}

The appropriate choice of reference frame for the correlation
function depends 
on our physical assumptions\footnote{Much to our chagrin, mild
assumption of the source are necessary even in the limit of infinite
statistics due to the loss of information in Eq. \ref{eq:corr} from
the 7 dimensional source function (8 minus the mass constraint) to the
6 dimensional correlation function.} 
of the source.  Traditionally
experimentalists and theorists in relativistic heavy ion collisions
have chosen the longitudinally comoving system (LCMS) as the
appropriate frame and deconvolve $\vec{q}$ into components
long, side and out.  The `long' component is that along the beam
direction; the `out' component is in the
direction of the average transverse momentum of the pair; and the
`side' component is the complementary orthogonal component,
perpendicular to the transverse motion of the pair.  This choice of
frame assumes
a longitudinally boost invariant source and has the very nice feature
that a static long lived source results in $\Ro \gg \Rs$.  This
feature is significant in searches for a quark gluon plasma where a
phase transition is accompanied by a large latent heat and long lived
source.

However, if the source is expanding, the LCMS parametrization sees a
Lorentz contracted outward radius ($\RoLCMS$), not the true coherence
length of the source in its rest frame ($\RoTRUE$).
As such, measurements of the HBT radii are distorted not only by the
fact that we measure only ``lengths of homogeneity'' but also by
Lorentz boosts of the source in the outward direction.  We can place
some boundaries on the actual size of the outward radius in the
source frame by noting
that the largest value of $\Ro$ is measured in the pair center of
mass frame (PCMS).  All other frames which are related to the PCMS
frame by a boost in the direction of the pair transverse momentum have a
smaller measured outward radius, $\RoPCMS/\gamma_{pr}$,
where $\gamma_{pr}$ is the gamma boost from the PCMS to the measured
frame.  A lower limit can also be placed on the true out
radius if we are willing to assume that the source is not moving
faster than the PCMS frame and is not going backward:
$\RoTRUE \ge \RoLCMS$.

Consider a simple example of a source moving at $\beta_{s} = 0.7$
toward the detector while emitting pairs with mean average momentum
of  $\langle k_T \rangle = 300$ MeV.  In this case, if the outward
radius measured in the LCMS frame is $\RoLCMS = 5$ fm, then the radius
measured in the PCMS is simply $\gamma_{pr} \RoLCMS = $ 11.8 fm.
Without a priori knowledge of the source velocity from, for example,
assumptions from the singles spectra, we can only limit the true source
to be between 5 and 11.8 fm.  However, the true source in its rest
frame has a radius of $\sqrt{\frac{1}{1-\beta_{s}}} \RoLCMS = $7 fm.

\section{Experimental Measures}

\subsection{The PHENIX Experiment}

\begin{figure}[experiment]
\vspace*{-.5cm}
\begin{center}%
\leavevmode
\epsfysize=6.5cm
\epsfbox{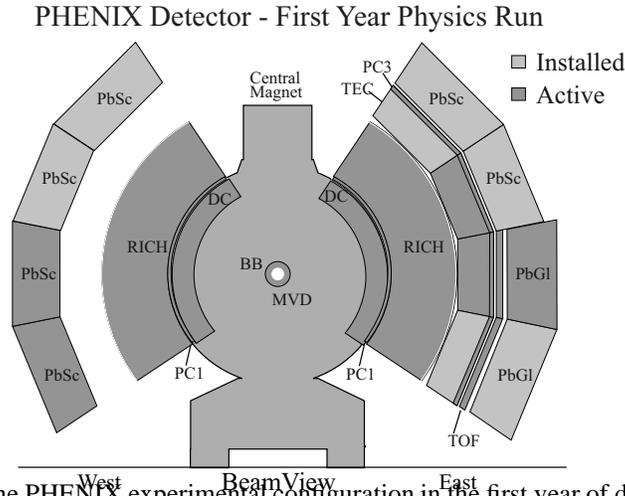}
\end{center}
\vspace*{-1cm}
\caption{The PHENIX experimental configuration in the first year of
data taking.}
\vspace*{-0.4cm}
\label{fig:experiment}
\end{figure}

The PHENIX experiment has been described in detail elsewhere \cite{PHENIX}.
It is composed of 4 ``arms'': two arms perpendicular to the beam 
specializing in electron, photon and hadron identification (``central arms'') 
and two arms for measuring muons (``muon arms'').  A beams-eye view of the 
central arms is shown in Figure \ref{fig:experiment}.

The central arms, covering a pseudo-rapidity region $|\eta|<0.35$, are
composed of a variety of subdetectors used in concert for particle
identification and momentum determination.  For the analysis in this
note, we use detectors only on the west arms.  The beam-beam counters
(BBC) and zero degree calorimeters (ZDC) were used to trigger on
inelastic interactions and determine the centrality of the collision;
the  drift chamber (DC), and pad chamber (PC) are used for three
dimensional pattern recognition and momentum determination; the
association of the track with a cluster in the electromagnetic
calorimeter (EMC) in conjunction with the BBC start time determines the
velocity of the particle \cite{mitchell}.

\subsection{Analysis Details}

Centrality is defined by the measured zero degree neutral energy in the ZDC
and the produced particle multiplicity measured in the BBC \cite{centrality}.
For this analysis we consider events in the 30\% most central range of 
the inelastic cross section.  The mean centrality of all pairs in this
sample is 10\%.  In the first year data sample this corresponds to
493K events after all offline cuts.

The mass of particles is determined in the analysis through a combination of 
momentum and velocity information.  Momentum dependent bands are both measured
experimentally and determined analytically from the momentum resolution of
the DC and PC ($\delta p / p = 0.6 \oplus 3.6\%p$) and timing resolution of 
the EMC-BBC pair (700 ps).  Pions are defined as being within 1.5$\sigma$ of
the pion mass peak and at least 2.5$\sigma$ from the kaon peak.  The resulting
dataset contains 3.1 million $\pi^+$ pairs and 3.3 million $\pi^-$ pairs.  The
mixed background is constructed from the pair dataset through the random
choice of two pions from pairs in different events.  Due to the changing detector 
acceptance as a function of collision position we require that all mixed
pairs come from events with a reconstructed BBC collision vertex within
1cm of each other.

To further minimize the influence of detector induced artifacts in the 
correlation function we introduce a number of cuts on both the real and mixed 
pair 
distributions.  We require that all pairs are at least 2cm from each other in 
the drift chamber to suppress close-track inefficiencies and ghosts.  Pairs 
that share the same EMC cluster are also removed from both the real and
mixed samples.

The correlation function is created by forming the ratio of real to 
mixed pairs versus the momentum difference of the pair.  This raw
correlation function is further corrected for the known Coulomb
interaction between the two pions.  The correction is performed by
making the simplifying assumption of a Gaussian source in the pair center
of mass frame and using an iterative procedure \cite{coulomb}.  The
extracted radii have been studied as a function of both the radius 
($R_{Coulomb}$) of the
source and the strength of the coulomb interaction ($\Lambda_{Coulomb}$) 
and are found to vary by no more
than .25 fm in any dimension versus reasonable variations of $R_{Coulomb}$ 
($\pm 20\%$)
and $\Lambda_{Coulomb}$ ($.5 \rightarrow 1.0$).  The results quoted below 
use the nominal extracted radius ($R_{Coulomb}=R_{inv}$) and full Coulomb 
correction ($\Lambda_{Coulomb} = 1.0$).

Finally, the correlation function is fit using a log-likelihood 
MINUIT based minimization method to the functional form
\[C_{2} = 1 + \lambda {\prod_{i=x,y...}} exp(-{R_{i}}^2 {q_{i}}^2 )\]
where $i$= inv for the one dimensional fit, $i$ = long, side, out
for the LCMS fit, etc.

Systematic errors enter primarily in the Coulomb correction and two
track cuts.  The total systematic errors of these effects is 8\% for
$\Rl$ and $\Rs$, and 4\% for $\Ro$ \cite{hbtprl}.

\subsection{Results}

\begin{figure}[htb]
\vspace*{-.5cm}
\begin{center}%
\leavevmode
\epsfysize=6.5cm
\epsfbox{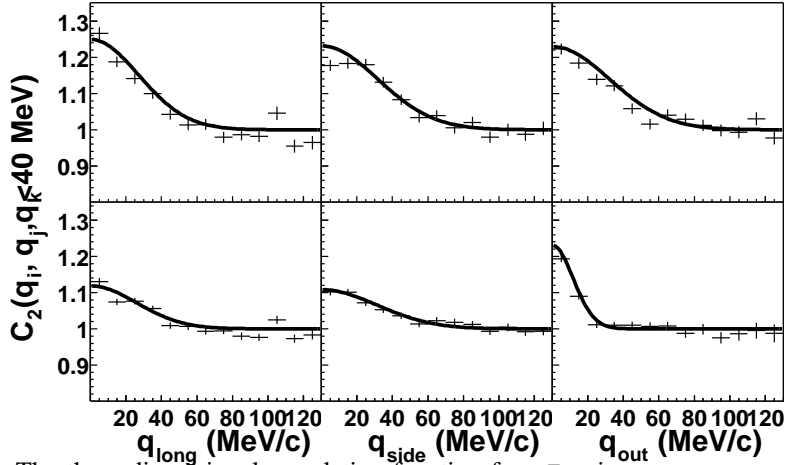}
\end{center}
\vspace*{-1cm}
\caption{The three dimensional correlation function for $\pi^-$ pairs
versus \ql, \qs, and \qo in both the LCMS frame (top)
and pair center-of-mass frame (bottom).  The data are plotted versus one
momentum difference variable while requiring the other two to be less
than 40 MeV/c.  The lines correspond to the fit to the entire distribution.}
\vspace*{-0.4cm}
\label{fig:correlation}
\end{figure}

A projection of the $\pi^-$ three dimensional correlation function is shown 
in Fig.~\ref{fig:correlation} with its corresponding fit.  This particular
plot includes the entire unbinned dataset.

The $\kT$ dependence of the correlation function provides extra information
on the dynamics of the source so we divide our results into three bins
in $\kT$ from 200-400 MeV, 400-550 MeV and 550-1000 MeV.  The corresponding
$\langle k_T \rangle$ are 333 MeV, 472 MeV and 633 MeV, respectively.  In each
$\kT$ bin we determine $\Rinv$, as well as three dimensional fits in the 
LCMS and PCMS frame.  The results are shown in Table \ref{t:table1}.

\begin{table} 
\caption{The \kT~dependencies of the $\pi^+$ and $\pi^-$ radii in
the LCMS and PCMS frames. All momenta are in MeV and
all radii are in fm.  The errors are statistical only. \label{t:table1}}
\begin{tabular}[]{ccccc}

 & $\kT$ (MeV)           & $200-400$ & $400-550$ & $550-1000$ \\

 & $\langle \kT \rangle$ & $333$     & $472$     & $633$      \\ \hline

 & 
$\Rinv$ &
$6.74 \pm 0.31$ &
$6.42 \pm 0.46$ &
$3.46 \pm 0.46$ \\

 & 
$\lambda_{LCMS}$   &
$0.423 \pm 0.037$ & 
$0.389 \pm 0.039$ & 
$0.287 \pm 0.048$ \\ 

$\pi^+$ &
$\Rl$ &
$6.01 \pm 0.45$ & 
$4.76 \pm 0.35$ & 
$2.97 \pm 0.38$ \\

 & 
$\Rs$ &
$4.81 \pm 0.30$ & 
$3.74 \pm 0.36$ & 
$2.79 \pm 0.37$ \\

 & 
$\Ro$ &
$4.78 \pm 0.30$ & 
$3.76 \pm 0.26$ & 
$2.59 \pm 0.46$ \\

 & 
$\RoPCMS$ &
$11.35 \pm 0.69$ &
$12.20 \pm 1.02$ &
$8.60 \pm 1.13$ \\ \hline

 & 
$\Rinv$ &
$6.00 \pm 0.30$ &
$5.96 \pm 0.41$ &
$4.58 \pm 0.48$ \\

 & 
$\lambda_{LCMS}$   &
$0.431 \pm 0.079$ & 
$0.405 \pm 0.067$ & 
$0.353 \pm 0.062$ \\ 

$\pi^-$ &
$\Rl$ &
$5.69 \pm 0.76$ & 
$4.77 \pm 0.49$ & 
$3.76 \pm 0.41$ \\

 & 
$\Rs$ &
$4.67 \pm 0.38$ & 
$4.13 \pm 0.45$ & 
$3.22 \pm 0.35$ \\

 & 
$\Ro$ &
$4.69 \pm 0.58$ & 
$3.75 \pm 0.40$ & 
$2.81 \pm 0.34$ \\ 

 & 
$\RoPCMS$ &
$11.27 \pm 0.72$ &
$12.42 \pm 1.18$ &
$11.89 \pm 1.73$ \\ 
\end{tabular} 
\end{table}

\section{Interpretation}

The recent correlation results  measured at RHIC, both by 
PHENIX \cite{hbtprl} and STAR \cite{starprl01} have generated
a great deal of interest and confusion in the theoretical community 
\cite{heinzHere}.  Much of the non perturbative measurements at 
RHIC in the first year of data taking can be well described in the context of 
a hydrodynamic model.  However, the correlation results have strained
the parameters of these models to unphysical values, prompting a theoretical
reevaluation.

\subsection{Comparisons}

\begin{figure}[htb]
\vspace*{-.5cm}
\begin{minipage}[t]{80mm}
\begin{center}%
\leavevmode
\epsfysize=10.0cm
\epsfbox{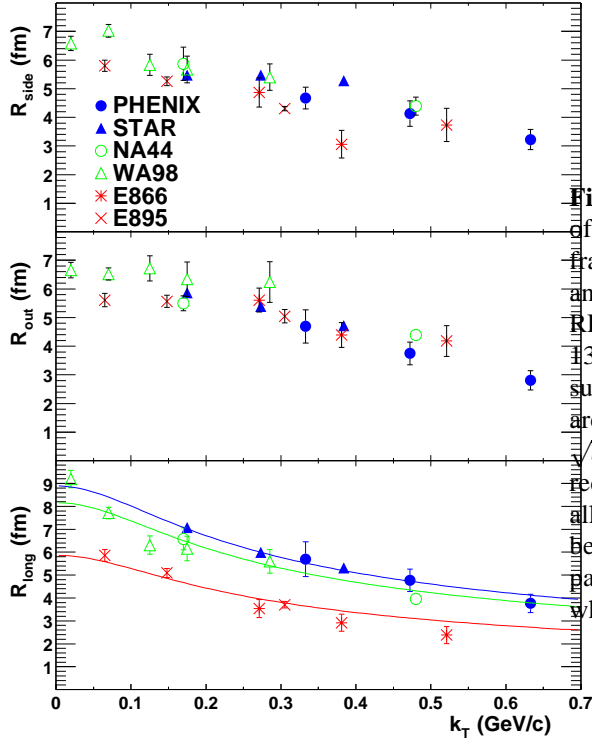}
\end{center}
\end{minipage}
\begin{flushright}
\begin{minipage}[t]{50mm}
\vspace*{-8cm}
\caption{The $\kT$ dependence of published radii in the LCMS frame
for various beam energies and experiments near mid-rapidity.  RHIC
measurements at $\snn = 130$GeV/c are in blue; SPS measurements at
$\snn = 17.3$GeV/c are in green; AGS results at 
$\snn = 4.9$ and $4.1$ GeV/c are in red.  The lines are a fit to $1/\sqrt{m_T}$ 
to all results ($\pi^+$ and $\pi^-$) for a given beam energy.  The data are
for $\pi^-$ pairs except for the NA44 results, which are for $\pi^+$.}
\label{fig:rvsroots}
\end{minipage}
\end{flushright}
\vspace*{-0.4cm}

\end{figure}

Figure \ref{fig:rvsroots} places the measurements by PHENIX
\cite{hbtprl} in the context of those at other energies and by other
experiments
\cite{starprl01,na44,wa98,soltz01,lis00}.\footnote{We plot the
\snn=4.9 \cite{soltz01} and \snn=4.1 \cite{lis00} together due to their 
minimal energy difference.}  As is clear, the 
\kT dependence of all three LCMS radii are very similar over a wide range
of beam energies ( $\snn=4.1$ to $130$ GeV/c).  Indeed the transverse
radii are nearly identical over the 
entire energy range.  The only clear energy dependence is in the longitudinal
component which increases moderately as a function of energy.

To quantify the differences in $\Rl$ as a function of beam energy, we fit each
$\kT$ dependence to a hydrodynamically inspired equation 
\cite{mak88,wie96}
\footnote{This functional form is supported by a theoretical 
approximation to first order of \Rl versus $T/\mT$.  In this form, 
$A=\tau_0 T$ where $\tau_0$ is the proper hadronization time.}
: $\Rl = A/\sqrt{\mT}$.  The results, overlaid
with the data in Fig.~\ref{fig:rvsroots}, are $A = 3.32 \pm 0.03$, 
$2.9 \pm 0.1$ and $2.19 \pm 0.05$~fm$\cdot$GeV$^\frac{1}{2}$ for $\snn$=130, 
17.3 and 4.9/4.1~GeV, respectively.

There are at least two outstanding issues that have avoided easy
physical descriptions: (1) the absolute magnitude of the radii and their
shape in $\kT$ at RHIC is strikingly similar to lower energy measurements
even though the particle multiplicity per unit rapidity increases by a factor
of three over the same energy range; (2) the ratio of the transverse radii 
$\Ro/\Rs$ is equal to 1 over all $\kT$.

The first puzzle, that of small radii, has been most troubling in the 
description of $\Rl$ which has been irreproducible in 
hydrodynamic calculations, with predictions 20-50\% higher than the data 
\cite{heinzHere}.

\begin{figure}[htb]
\vspace*{-.5cm}
\begin{minipage}[t]{80mm}
\begin{center}%
\leavevmode
\epsfysize=9.0cm
\epsfbox{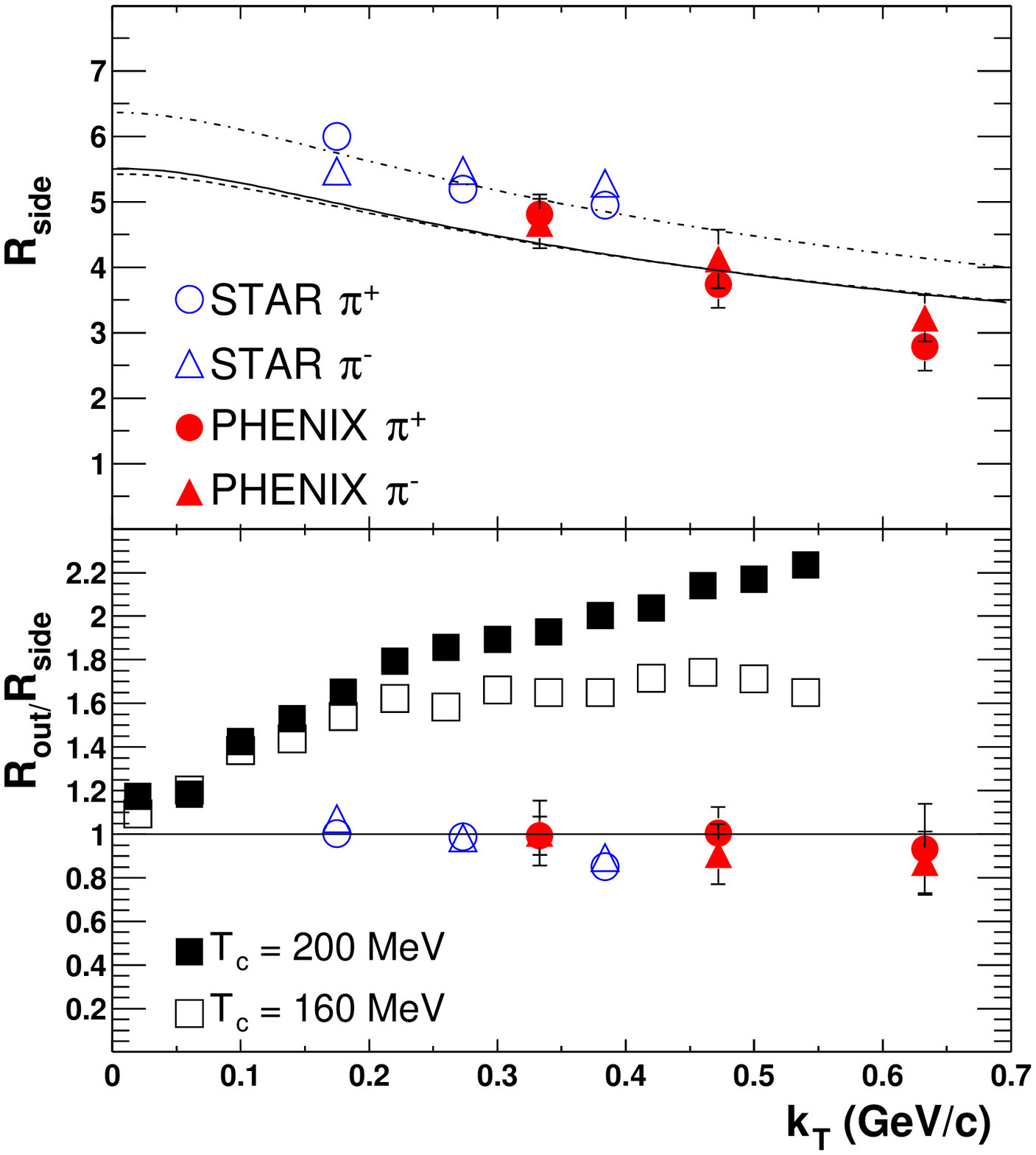}
\end{center}
\end{minipage}
\begin{flushright}
\begin{minipage}[t]{55mm}
\vspace*{-8cm}
\caption{The top panel shows the measured \Rs from identical pions for
both STAR and PHENIX.  The solid line is a fit of Eq.~\ref{eq:second} to the PHENIX data,
and the dashed line is the same fit for Eq.~\ref{eq:first}.  The dot-dashed line
is a fit of Eq.~\ref{eq:second} to the STAR data.  The bottom panel shows the ratio 
\Ro/\Rs as a function of \kT overlaid with theoretical predictions for a
phase transition for two critical temperatures.}
\label{fig:rsandro}
\end{minipage}
\end{flushright}
\vspace*{-0.4cm}
\end{figure}

The second puzzle, regarding the $\Ro/\Rs$ ratio, is quantified in Figure 
\ref{fig:rsandro} where we plot this ratio versus $\kT$ for both experiments 
at RHIC overlaid with theoretical predictions of a model that includes 
both a phase transition and a period of hadronic rescattering calculated
with a microscopic transport code \cite{soff}.  The model includes
two scenarios with different critical temperatures and the authors
note that a higher critical temperature leads to a longer period of
rescattering in the hadronic phase.  This in turn leads to a larger
$\Ro/\Rs$ ratio.
The data are completely inconsistent with this theoretical prediction, and
with any physically reasonable hydrodynamic theoretical calculation to
date. \cite{heinzHere,rischke}

\subsection{Hydrodynamical Fits}

The field of correlation measurements has adopted a
method of interpretation based upon fits of the $\kT$ dependence of the
radii.  The effect of Bose-Einstein correlations on a boost-invariant,
hydrodynamically expanding source can be calculated analytically
\cite{chapman95} to first order

\begin{equation}
\Rs^2 = \frac{{R_{geom}}^2}{1+\beta_f^2(\frac{m_T}{T})}
\label{eq:first}
\end{equation}
where $R_{geom}$ is the geometric radius of the source, $m_T^2 =
k_T^2+m_\pi^2$, $\beta_f$
is the boost velocity ($\beta_T = \beta_f \frac{\rho}{R_{geom}}$) and
$T$ is the temperature.  A fit of this functional form to the PHENIX
data assuming T=125 MeV and $\beta_T = 0.69$ \cite{hbtprl,jane}
\footnote{These values are taken from fits to the singles spectra for
the 5\%-15\% centrality bin.} is
shown as a dashed line in Figure~\ref{fig:rsandro}.  The extracted
radius is $R_{geom}=$6.7$\pm$0.2 fm.

Adding a second term in the expansion leads to the form \cite{wiedemann96}

\begin{equation}
\Rs^2 = \frac{{R_{geom}}^2}{1+\eta_f^2(\frac{1}{2}+\frac{m_T}{T})}
\label{eq:second}
\end{equation}
in which $\eta_f$ is the transverse rapidity boost 
$\eta_T = \eta_f \frac{\rho}{R_{geom}}$.  The result of a fit to the
PHENIX data with this form yields $R_{geom} =$8.1$\pm$0.3 fm.  A fit
of the STAR data to equation \ref{eq:second} yields 
$R_{geom}=$9.4$\pm$0.1 fm.  These two fits are also shown in
figure~\ref{fig:rsandro}  All of these values for the geometric radius
are much larger than the comparable 1D rms radius of the Au nucleus.

Taken together, the STAR and PHENIX data imply a steeper slope of
$\Rs$ vs. $k_T$ then the fit, suggesting a higher $\beta/T$ ratio.
However, we should reserve our judgment of this issue until a full
systematic study can be performed by both experiments with the larger
second year data sets.  At present, systematic errors in the
experimental measures can completely account for the difference
between the experiments.

\section{Conclusion}

We have presented the first year measurements of identical pion
correlations as a function of $k_T$ by the PHENIX experiment at
$\snn$=130 GeV.  Comparisons to lower energies reveals a mild energy
dependence in the longitudinal direction with no discernible
transverse radius dependence.  The ratio $\Ro/\Rs$ is consistent with
unity for all $k_T$ in contrast to theoretical predictions that
include a phase transition.

 
\section*{Acknowledgments}

The author would like to thank Dr. R.A.~Soltz for enlightening
discussions and unadulterated support.

This work was performed under the auspices of the U.S. Department of
Energy by the University of California, Lawrence Livermore National
Laboratory under Contract No. W-7405-Eng-48.


\vfill\eject
\end{document}